\documentstyle[aps,twocolumn]{revtex}
\begin{document}
\tightenlines
\title{Decoherence of gauge-invariant metric fluctuations
during inflation}
\author{Mauricio Bellini\footnote{E-mail address: mbellini@mdp.edu.ar}}
\address{Instituto de F\'{\i}sica y Matem\'aticas, \\
Universidad Michoacana de San Nicol\'as de Hidalgo,\\
AP: 2-82, (58041) Morelia, Michoac\'an, M\'exico}
\maketitle
\begin{abstract}
I study the gauge-invariant fluctuations of the metric during
inflation. In the infrared sector the metric fluctuations can be
represented by a coarse-grained field. We can write a Schr\"odinger
equation for the coarse-grained metric fluctuations which is related
to an effective Hamiltonian for a time dependent parameter of mass
harmonic oscillator with a stochastic external force.
I study the wave function for a power-law expanding universe.
I find that the phase space of the quantum state for super Hubble
scalar metric perturbations loses its coherence at the end of inflation.
This effect is a consequence of interference between the super
Hubble metric perturbations and its canonical conjugate variable,
which is produced by the interaction of
the coarse-grained scalar metric fluctuation
with the environment.
\end{abstract}
\vskip 2cm
\noindent
PACS numbers: 98.80.Cq, 04.62.+v\\
\vskip .5cm

The inflationary model\cite{linde90}
solves several difficulties which arise from the standard cosmological model,
such as the horizon, flatness and monopole problems, and it
provides a mechanism for the creation of primordial density of fluctuations,
nedeed to explain the structure formation\cite{starobinsky}.
The most
widely accepted approach assumes that the inflationary phase is driving by a
quantum scalar field $\varphi $ associated to a scalar potential $V(\varphi)$.
Within this
perspective, the stochastic inflation proposes to describe the dynamics of
this quantum field on the basis of two pieces: the homogeneous
and inhomogeneous components. Usually the homogeneous one is interpreted
as a classical field $\phi_c(t) $
that arises from the vacuum expectation value of
the quantum field. The inhomogeneous component $\phi(\vec x,t)$
are the quantum fluctuations.
The field that take into account only the modes with wavelengths
larger than the now observable universe is called coarse-grained field
and its dynamics is described by a second order
stochastic equation\cite{BCMS}.

Stochastic inflation has played an important role in inflationary cosmology
in the last two decades. This approach gives the possibility to
make a description of the matter field fluctuations in the infrared (IR)
sector by means of the coarse-grained matter field\cite{starobinsky,Habib}.
Since these perturbations are classical on super Hubble scales,
in this sector one can make a standard stochastic
treatment for the coarse-grained matter field. The IR sector is
very important because the spatially inhomogeneities in
super Hubble inflationary scales would explain the present day
observed matter structure in the universe.
Matter field fluctuations are responsible for metric fluctuations
around the background Friedmann-Robertson-Walker (FRW)
metric. When
these metric fluctuations do not depend
on the gauge, the perturbed globally flat
isotropic and homogeneous universe is described by\cite{3}
\begin{equation}\label{m}
ds^2 = (1+2\psi) \  dt^2 - a^2(t) (1-2\chi) \  dx^2,
\end{equation}
where $a$ is the scale factor of the universe and ($\psi$, $\chi$) are
the perturbations of the metric.
In the particular
case where the tensor $T_{\alpha\beta}$ is diagonal, one obtains:
$\chi = \psi$\cite{4a}.
I consider a semiclassical expansion for the
scalar field $\varphi(\vec x,t) = \phi_c(t) + \phi(\vec x,t)$\cite{BCMS},
with expectation values
$\left<0|\varphi|0\right> = \phi_c(t)$ and $\left<0|\phi|0\right>=0$. Here,
$\left.|0\right>$ is the vacuum state.
Due to $\left<0|\chi|0\right> =0$, the expectation
value of the metric (\ref{m}) gives the background
metric that describes a flat FRW spacetime.

The aim of this paper is the study of decoherence in the phase space
for the quantum state that describes the scalar metric perturbations
on super Hubble scales during inflation. This issue have been studied
in \cite{a,b} by means of the Wigner function. In this work I study this
tool using the Schr\"odinger wave function for the IR gauge-invariant
scalar metric fluctuations.

The Einstein's equations can be linearized with respect to
$\phi$ and $\chi$\cite{6}
\begin{eqnarray}
 \ddot\chi &+& \left(\frac{\dot a}{a}
- 2 \frac{\ddot\phi_c}{\dot\phi_c} \right)
\dot \chi - \frac{1}{a^2} \nabla^2 \chi \nonumber \\
&+& 2\left[
\frac{\ddot a}{a} - \left(\frac{\dot a}{a}\right)^2 - \frac{\dot a}{a}
\frac{\ddot\phi_c}{\dot\phi_c}\right] \chi =0, \label{1}\\
\frac{1}{a}& \frac{d}{dt}& \left( a \chi \right)_{,\beta} =
\frac{4\pi}{M^2_p} \left(\dot\phi_c \phi\right)_{,\beta} , \\
\ddot\phi& +& 3 \frac{\dot a}{a} \dot\phi -
\frac{1}{a^2} \nabla^2 \phi + V''(\phi_c) \phi
+  2 V'(\phi_c) \chi- 4 \dot\phi_c \dot\chi =0,
\end{eqnarray}
where $\beta = 0,1,2,3$.
The dynamics of $\phi_c$ being described by the equations
\begin{eqnarray}
&& \ddot\phi_c + 3 \frac{\dot a}{a} \dot\phi_c + V'(\phi_c) = 0, \\
&& \dot \phi_c = - \frac{M^2_p}{4 \pi} H'_c(\phi_c).
\end{eqnarray}
Here, the prime denotes the derivative with respect to $\phi_c$ and
$H_c(\phi_c) \equiv {\dot a\over a}$. As in a previous
work\cite{pr2000}, we can make the transformation
$h =
e^{1/2 \int \left({\dot a \over a}-
{2 \ddot\phi_c \over \dot\phi_c} \right) dt} \  \chi$
in eq. (\ref{1})
\begin{eqnarray}
\ddot h &-& \frac{1}{a^2} \nabla^2 h
 \nonumber \\
&-& \left[ \frac{1}{4}
\left(\frac{\dot a}{a} - 2 \frac{\ddot \phi_c}{\dot\phi_c} \right)^2
+ \frac{1}{2} \left( \frac{\ddot{a} a - \dot{a}^2}{a^2}
- \frac{2 \frac{d}{dt}\left(\ddot\phi_c \dot\phi_c\right)
- 4 \ddot\phi^2_c}{\dot\phi^2_c} \right)\right.\nonumber\\
& +& 2\left( \frac{\ddot a}{a} - \left( \frac{\dot a}{a}\right)^2
- \left.
\frac{\dot a}{a} \frac{\ddot\phi_c}{\dot\phi_c} \right) \right] h
= 0\label{2}.
\end{eqnarray}
The eq. (\ref{2}) is a Klein - Gordon like
equation for the
redefined fluctuations of the metric $h(\vec x,t)$.
The term
\begin{eqnarray}
\mu^2(t) &=&
\frac{1}{4}
\left(\frac{\dot a}{a} - 2 \frac{\ddot \phi_c}{\dot\phi_c} \right)^2
+ \frac{1}{2} \left( \frac{\ddot{a} a - \dot{a}^2}{a^2}
- \frac{2 \frac{d}{dt}\left(\ddot\phi_c \dot\phi_c\right)
- 4 \ddot\phi^2_c}{\dot\phi^2_c} \right) \nonumber \\
&-& 2\left( \frac{\ddot a}{a} - \left( \frac{\dot a}{a}\right)^2
-\frac{\dot a}{a} \frac{\ddot\phi_c}{\dot\phi_c} \right) ,
\end{eqnarray}
plays the role of time dependent squared mass in the eq.
(\ref{2}), and also can be written as $\mu^2(t) ={k^2_0\over a^2(t)}$.
Here, $k_0(t)$ is
the time dependent wavenumber that separates the infrared (IR)
and ultraviolet (UV) sectors.

Now I writtes the field $h$ as a Fourier expansion in terms of the
modes $h_{\vec k} =e^{i \vec k. \vec x} \xi_k(t)$
\begin{equation}
h(\vec x,t) = \frac{1}{(2\pi)^{3/2}} \int
d^3 k \left[ a_k h_k + a^{\dagger}_k h^*_k\right],
\end{equation}
where $a_k$ and $a^{\dagger}_k$ are the annihilation and creation
operators with commutation relations $[a_{\vec k},a^{\dagger}_{\vec k'}] =
\delta^{(3)}(\vec k - \vec k')$ (the asterisk denotes the complex
conjugate).
The matter field perturbations, written as a Fourier expansion, are
\begin{equation}
\phi(\vec x,t) = \frac{1}{(2\pi)^{3/2}} \int
d^3 k \left[ a_k \phi_k + a^{\dagger}_k \phi^*_k\right],
\end{equation}
where $\phi_{\vec k} =e^{i \vec k. \vec x} u_k(t)$.
Since $\chi = \psi$, hence the metric and matter perturbations
are anticorrelated outside the horizon: $\xi_k
=- \phi_c(t) \  e^{-1/2 \int \left({\dot a \over a}-
{2 \ddot\phi_c \over \dot\phi_c} \right) dt} \  u_k$\cite{4a}.
The equation for the time dependent modes $\xi_k(t)$
is\cite{pr2000}
\begin{equation}\label{tm}
\ddot\xi_k + \omega^2_k(t) \xi_k =0,
\end{equation}
where $\omega_k(t)$ is
the time dependent frequency for each mode with wavenumber $k$:
$\omega^2_k(t) = [k^2/a^2 - \mu^2(t)]$.
The commutation relation for $h$ and $\dot h$ is
$[h(\vec x,t),\dot h(\vec x',t)]= {\rm i}
\delta^{(3)} (\vec x - \vec x')$ for
$\xi_k \dot\xi^*_k - \dot\xi_k \xi^*_k ={\rm i}$.
The modes $\xi_k(t)$ are real on super Hubble scales, so that they
hold $\xi_k \dot\xi^*_k - \dot\xi_k \xi^*_k =0$. This fact is
due to the classicality of $h$ for cosmological scales.

Now I consider the
field $h(\vec x,t)$ on the IR sector. For $k \ll k_0(t)$
the coarse - grained field $h_{cg}(\vec x,t)$ can be written as
\begin{equation}\label{s}
h_{cg}(\vec x,t) = 
\frac{1}{(2\pi)^{3/2}} \int d^3 k \  \theta(\epsilon k_0-k)
[a_k h_k + a^{\dagger}_k h^*_k],
\end{equation}
where $\epsilon \ll 1$ is a dimensionless constant
and $\theta(\epsilon k_0-k)$ is the Heaviside function which takes into
account only the modes on cosmological scales.

{\em Stochastic approach and Heisenberg representation:}
Replacing
the eq. (\ref{s}) in (\ref{2}), one obtains the following
stochastic equation for $h_{cg}$
\begin{equation}\label{st}
\ddot h_{cg} -\frac{k^2_0}{a^2} h_{cg} = \epsilon \left[
\frac{d}{dt} \left(\dot k_0 \eta \right) + 2 \dot k_0 \kappa \right],
\end{equation}
where the noises $\eta$ and $\kappa$ are
\begin{eqnarray}
\eta(\vec x,t) &=& \frac{1}{(2\pi)^{3/2}} \int d^3 k \  \delta
(\epsilon k_0 - k)
[a_k h_k + a^{\dagger}_k h^*_k ], \\
\kappa(\vec x,t) &=& \frac{1}{(2\pi)^{3/2}}
\int d^3 k \  \delta(\epsilon k_0 - k)
[a_k \dot h_k + a^{\dagger}_k \dot h^*_k ].
\end{eqnarray}
The eq. (\ref{st}) can be written as
\begin{equation}\label{a7}
\ddot h_{cg}- \left[\frac{k_0(t)}{a(t)}\right]^2 \  h_{cg}+
\xi_c(\vec x, t)=0,
\end{equation}
where
$\xi_c(\vec x, t)=- \epsilon \left[{d \over dt}(\dot k_0 \eta^{(c)})
+ 2 \dot k_0 \kappa \right]$.
This effective noise arises from the modes of the short wavelength sector
that crosses the Hubble horizon towards the IR sector.
This effect generates the growth of the number of
degrees of freedom in the IR sector.
The eq. (\ref{a7}) is related with the effective Hamiltonian
\begin{equation}\label{a8}
H_{eff}(h_{cg},t)= \frac{1}{2} P^2_{cg}+
\frac{1}{2} \omega^2_k(t) \  \left(h_{cg}\right)^2
+ \xi_c h_{cg},
\end{equation}
where $P_{cg}\equiv \dot h_{cg}$ and $\omega^2_k
\simeq -\mu^2(t)={k^2_0\over a^2}$ for $k^2 \ll k^2_0$.
Note that $\xi_c $
plays the role of an external classical stochastic
force in the effective Hamiltonian
(\ref{a8}) and $\mu^2(t)$ is a time dependent squared mass which
depends of the particular cosmological model which we consider.
With respect to the Hamiltonian (\ref{a8}), one
can write the Schr\"odinger equation
\begin{eqnarray}
{\rm i}\frac{\partial }{\partial t} \Psi(h_{cg},t) &=&
-\frac{1}{2} \frac{\partial^2}{\partial h^2_{cg}}
\Psi(h_{cg},t) \nonumber \\
&+& \left[
\xi_c \  h_{cg}-\frac{1}{2} \  \mu^2(t) \left(h_{cg}\right)^2 \right] 
\Psi(h_{cg},t),\label{scho}
\end{eqnarray}
where $\Psi(h_{cg},t)$ is the wave function that
characterize the system for the $h_{cg}$-representation.
Since generally $\mu(t)$ depends on time, hence the
Hamiltonian (\ref{a8}) is non - conservative, also in the case when
one would neglects the stochastic force $\xi_c$. The only case where
$\mu$ is a constant is for a de Sitter expansion.
The wave function
for the system being given by\cite{10}
\begin{eqnarray}
\Psi(h_{cg},t) &=& \frac{1}{(2\pi)^{1/4} \Delta^{1/2}}
e^{-\frac{1}{4\Delta^2} \left[h_{cg} - h_{cl}\right]^2} e^{i \frac{h^2_{cg}}{
\Delta^2} \left[2\frac{\dot B}{B} \Delta^2 + \frac{{\cal R}(t)}{\sigma^2}
\right]} \nonumber \\
&\times & e^{i \frac{h_{cg}}{\Delta^2} \left[ \Delta^2 \left(
P_{cl} - \frac{\dot B}{B} h_{cl}\right) - \frac{{\cal R}(t) h_{cl}}{
2\sigma^2}\right] } e^{i \gamma(t)},
\end{eqnarray}
where $\gamma(t)$ is an arbitrary phase and ($B(t_0)=1, \dot B(t_0)=0$)
are the initial conditions.
The function $B(t)$ is the solution
of the equation
\begin{equation}\label{equa}
\ddot B - \mu^2(t) B =0,
\end{equation}
where $\mu^2(t) = {k^2_0 \over a^2}$. I denote $P_{cl} =
\left< \dot h_{cg}\right>$ and $h_{cl} =
\left< h_{cg}\right>$. Furthermore, the parameter $\Delta(t)$ is given
by
\begin{equation}
\Delta^2(t) = \frac{B^2(t)}{\sigma^2} \left[\sigma^4 + {\cal R}^2(t)
\right],
\end{equation}
for
\begin{equation}
{\cal R}(t) = {\Large\int}^t dt' \  \frac{1}{2 B^2(t')},
\end{equation}
and $\sigma^2 =  \Delta^2(t_0)$.
The expectation value for the effective energy is
\begin{equation}
\left<E_{eff}\right> = E_{cl} + \frac{1}{8 \Delta^2} -
\frac{\mu^2(t)}{2} \Delta^2 + {\cal D}(t),
\end{equation}
where
\begin{equation}
E_{cl} = \frac{P^2_{cl}}{2} - \frac{\mu^2(t)}{2} h^2_{cl} +
\xi_c h_{cl},
\end{equation}
and ${\cal D}(t)={1\over 2} \left({\dot B\over B} \Delta +
{{\cal R}\over 2 \Delta \sigma^2} \right)^2$ is the function
which takes into account decoherence of the phase space.
If ${\cal D}(t)$ increases with time, hence the phase space
$(h_{cg},P_{cg})$ will be decoherentized.

Finally, the squared fluctuations $\left< h^2_{cg}\right>$
and $\left<P^2_{cg}\right>$ are
\begin{eqnarray}
\left< h^2_{cg}\right> & = & h^2_{cl}(t) + \Delta^2(t), \label{aa1}\\
\left< P^2_{cg} \right> & = & P^2_{cl}(t) +
\frac{1}{4 \Delta^2(t)} + 2 {\cal D}(t) \label{aa2},
\end{eqnarray}
where $\left<P_{cg}\right> = P_{cl}$ and $\left<h_{cg}\right>=h_{cl}$.
Note that we are dealing with $h_{cg}$-representation, so that
$P_{cg} \equiv - i {\partial \over \partial h_{cg}}$.
The second terms in eqs. (\ref{aa1}) and (\ref{aa2}) represent the 
quantum fluctuations which depend only onthe choice of $\mu^2(t)$. In 
particular $\mu^2(t)$ being given by the cosmological model we
are considering. Furthermore, the third term in eq. (\ref{aa2}) represents
quantum fluctuations related to decoherence in the phase space. In the case
of a pure quantum state (i.e., in a coherent state) the function
${\cal D}(t)$ becomes zero.

{\em An example: power-law expanding universe.}
Now we can study the case of a power-law expansion for the universe.
In such an expanding universe the scale factor evolves
as $a \propto \left(t/t_0\right)^p$, and the Hubble parameter
is $H(t) = p/t$. Furthermore, the classical background field is
$\phi_{cl}(t) = \phi^{(0)}_{cl} - m \  {\rm ln}\left[(t/t_0)p\right]$,
where $m$ is the mass of the inflaton field.
The squared parameter of mass $\mu^2(t)$ is given by
\begin{equation}
\mu^2(t) = \frac{1}{4} \left(p^2 +4\right) t^{-2}.
\end{equation}
The equation (\ref{equa}) for $B(t)$ takes the form
\begin{equation}
\ddot B(t) - \frac{1}{4} \left(p^2 +4\right) B(t) = 0,
\end{equation}
which has the general solution
\begin{equation}
B(t) =
c_1 \  t^{\frac{1}{2} \left(1+ \sqrt{p^2+5}\right)} 
+ c_2 \  t^{\frac{1}{2} \left(1- \sqrt{p^2+5}\right)},
\end{equation}
where $c_1$ and $c_2$ are arbitrary constants.
From the general solution for $B(t)$ one obtains 
\begin{eqnarray}
&& B(t_0) = c_1+c_2, \\
&& \dot B(t_0) = \frac{1}{2}\left[c_1+c_2+\sqrt{p^2+5}(c_1-c_2)\right]=0.
\end{eqnarray}
Hence, the initial conditions $B(t_0)=1$ and $\dot B(t_0)=0$ implies:
\begin{equation}
c_1= \frac{\sqrt{p^2+5}-1}{2\sqrt{p^2+5}}, \qquad c_2=
\frac{1+\sqrt{p^2+5}}{2\sqrt{p^2+5}}.
\end{equation}
This
means that, for $p>3.04$ needed to inflation takes place, 
the asymptotical behaviour of the functions ${\cal R}(t)$,
$\Delta^2(t)$ and ${\cal D}(t)$, are
\begin{eqnarray}
\left.{\cal R}(t)\right|_{t\gg 1} & \simeq & 
\frac{\sqrt{p^2+5}}{\left(\sqrt{p^2+5}-1\right)
\left(1+\sqrt{p^2+5}\right)} \nonumber \\  
&\times & {\rm ln}\left[\frac{2\sqrt{p^2+5}}{\sqrt{p^2+5}-1}\right],\\
\left.\Delta^2(t)\right|_{t\gg 1} & \propto &
t^{1+\sqrt{p^2+5}},\\
\left.{\cal D}(t)\right|_{t\gg 1} & \propto & 
t^{\frac{1}{2} \left(\sqrt{p^2+5}-1\right)}.\label{D}
\end{eqnarray}
Note that decoherence ${\cal D}(t)$ increases as $
t^{{1\over 2} \left(\sqrt{p^2+5}-1\right)}$.
This fact, and large squeezing of quantum state,
lead to an effective quantum-to-classical
transition of the phase space, which describes the gauge-invariant metric
fluctuations on cosmological scales.
The density of probability to find the universe (in the IR sector) with a
given value of field $h_{cg}$, in a given time $t$ is
(the asterisk denotes the complex conjugate)
\begin{eqnarray}
P(h_{cg},t) & = & \Psi(h_{cg},P_{cg}) \  \Psi^*(h_{cg},P_{cg})
\nonumber \\
&=&
\frac{1}{(2\pi)^{1/2} \Delta} e^{-\frac{1}{2 \Delta^2}
\left[h_{cg} - h_{cl}\right]^2},
\end{eqnarray}
which, for late times gives the asymptotic expression
\begin{eqnarray}
\left. P(h_{cg},t)\right|_{t\gg 1} 
& \simeq & \frac{1}{(2\pi)^{1/2} t^{\frac{1}{2}
\left(1+\sqrt{p^2+5}\right)}} \nonumber \\
&\times &  \exp{\left[-\frac{1}{t^{1+\sqrt{p^2+5}}}
\left[ h_{cg} - h_{cl}\right]^2\right]}.
\end{eqnarray}

Note that $P(h_{cg},t)$ is peacked at $h_{cl}$, 
but
the squared dispersion grows monotonically as $t^{1+\sqrt{p^2+5}}$. 

{\em General comments:}
In this work I shown that the gauge-invariant
IR metric fluctuations can
be described by means of the Schr\"odinger equation. In the
power-law expanding model here studied, I find that the
wave function is peacked in $h_{cg} = h_{cl}$, but the dispersion
increases monotonically with time. 
It is a consequence of quantum
mechanical fluctuations. However,
since ${\cal D}(t)$ increases with time, also the squared fluctuations
$\left<P^2_{cg}\right>$ grows during power-law inflation. Notice we are
dealing with $h_{cg}$-representation and the late times quantum mechanical
fluctuations $\left<P_{cg}\right>$ should be decreasing in a pure quantum
state. However, due to the interference between the fluctuations
of $h_{cg}$ and $P_{cg}$ (i.e., due to decoherence in the phase space),
the squared fluctuations $\left<P^2_{cg}\right>$ increase with time
[see. eqs. (\ref{aa2}) and (\ref{D})]. 
In the case here studied, the origin of decoherence resides in the
interchange of degrees of freedom between the IR and UV sectors.
So, the coarse-grained metric fluctuations $h_{cg}$ interacts with
the environment that takes into account only short-wavelength modes.
Hence, the effective Hamiltonian (\ref{a8}) describes an open
quantum system 
with a temporal dependence of the effective parameter of mass $\mu(t)$, which
is originated in the self-interaction of metric fluctuations. Furthermore,
such $h$-self-interaction is a consequence of the self-interaction
of the inflaton field in a power-law expanding universe.
Due to the quantum-to-classical transition of super Hubble
metric fluctuations here demonstrated, it is possible (and thus equivalent)
to describe such fluctuations by means of
a stochastic description. In such a description the interaction between
the coase-grained and sub Hubble metric perturbations is represented
by a Gaussian and white noise and the dynamics for the transition
probability being given by the Fokker-Planck equation.
(The reader can see, for example, a stochastic treatment---but
for matter field fluctuations---in ref. \cite{BCMS}.)

\end{document}